\documentclass[twocolumn,showpacs,preprintnumbers,amsmath,amssymb,fleqn]{revtex4} 
\usepackage{amsmath,amssymb,graphicx}
\begin{document}

\title{Simultaneous cooling of degenerate mechanical modes in unresolved sideband regime via optical and mechanical nonlinearities}
\author{Shuang-Shuo Chu$^{1}$}
\author{Han-Qiu Zhang$^{1}$}
\author{Jian-Song Zhang$^{1}$}
\email{jszhang1981@zju.edu.cn}
\author{Wen-Xue Zhong$^{1}$}
\author{Guang-Ling Cheng$^{1}$}
\author{Ai-Xi Chen$^{2}$}
\email{aixichen@zstu.edu.cn}
\affiliation{$^{1}$Department of Applied Physics, East China Jiaotong University,
Nanchang 330013, People's Republic of China \\
$^{2}$ Department of Physics, Zhejiang Sci-Tech University, Hangzhou 310018, People's Republic of China}

\begin{abstract}
We propose a scheme to simultaneously cool multiple degenerate mechanical modes in optomechanical systems beyond the resolved sideband regime.
In general, one of the main obstacles for cooling degenerate mechanical modes is the so-called dark-mode effect.
The Duffing nonlinearities (mechanical nonlinearities) can be used to overcome the dark-mode effect of degenerate mechanical modes.
A second-order nonlinear medium (optical nonlinearity) is introduced to accomplish the ground-state cooling of degenerate mechanical modes
beyond the resolved sideband regime. We find the dark mode of degenerate mechanical modes can be broken when the mechanical nonlinearities of different
mechanical modes are not very close. Our scheme paves the way toward the implementation of simultaneous ground-state cooling of degenerate mechanical modes
of optomechanical systems beyond the resolved sideband regime in experiments.

\end{abstract}
\maketitle

\section{Introduction}
In recent years, optomechanical systems have received a lot of attention since they have many applications
including highly sensitive measurement of tiny displacement, creation of nonclassical states
of light or mechanical motion, and quantum information processing \cite{Agarwal2013,Aspelmeyer2014, Bowen2015,Lv20151,Lv20152,Zhang2019,Zhang20201,Lv2021}.
Of particular interest are multimode optomechanical systems with two or more mechanical oscillators \cite{Mancini2002,Borkje2011,Stannigel2012,Massel2012,Spethmann2016,Piergentili2018,Riedinger2018,Ockeloen2018,Yang2020,Kotler2021,Lepinay2021,Bhattacharya2008,Xu2013,Mercade2021,Lai2021}.
Multimode optomechanical systems have a wide range of applications such as the generation of entanglement between two or more mechanical oscillators,
the study of quantum many-body effects, and highly sensitive sensors \cite{Agarwal2013,Aspelmeyer2014, Bowen2015}.
In the applications of multimode optomechanical systems, the simultaneous ground-state cooling of multiple degenerate mechanical resonators is
indispensable \cite{Schwab2005}.

Unfortunately, there are two main obstacles for the simultaneous ground-state cooling of several degenerate mechanical modes in standard sideband cooling.
One is the resolved sideband condition, i.e., the decay rate of an optical cavity must be smaller than the frequencies of mechanical modes \cite{Bowen2015}.
This condition typically requires that the finesse of the optical cavity should be very high and limits the size of the mechanical resonators to be cooled.
This restriction can be overcome with the help of an auxiliary mechanical mode \cite{Ojanen2014} or a coherent auxiliary cavity \cite{Liu2015}.
The other main obstacle is the so-called dark-mode effect, which suppresses the ground-state cooling of degenerate mechanical modes significantly.
This effect appears if two or more degenerate mechanical modes couple to one common optical cavity mode \cite{Genes2008,Dong2012,Wang2012,Shkarin2014,Kuzyk2017}. Physically, a dark mode formed by two degenerate mechanical modes is decoupled from the cavity mode of the system completely. Thus it is very difficult to extract thermal excitations of the dark mode through the cooling channel of the optical mode. Later, the dark-mode effect was demonstrated experimentally in \cite{Ockeloen2019}. Up to now, several schemes were proposed for
cooling multiple mechanical oscillators simultaneously \cite{Lai2018,Lai20212,Lai2020,Naseem2021,Sommer2019,Huang20221}. In Ref. \cite{Lai2018}, the authors have pointed out that
the ground-state cooling of several mechanical oscillators could be accomplished in the resolved-sideband regime by introducing an optomechanical interface. It was shown that the dark modes can be broken by introducing a phase-dependent phonon-exchange interaction \cite{Lai2020}. The quantum reservoir engineering method was also used to realize ground-state cooling of several mechanical resonators \cite{Naseem2021}.
The authors of Ref. \cite{Sommer2019} suggested the thermal energy from many mechanical modes within a large frequency bandwidth can be extracted with the help of a standard cold-damping technique.
Very recently, an auxiliary cavity mode was introduced to overcome the dark-mode effect of multiple degenerate mechanical modes \cite{Huang20221,Huang20222}.
Consequently, the simultaneous ground-state cooling of degenerate mechanical modes can be realized in multimode optomechanical systems in the resolved sideband regime \cite{Huang20221}.

In the present work, we propose a scheme to cool several degenerate mechanical oscillators simultaneously in the unresolved sideband regime
with the help of optical and mechanical nonlinearities. In order to break the dark modes formed by degenerate mechanical modes,
we introduce the Duffing nonlinearities (mechanical nonlinearities). Note that the Duffing nonlinearities of different mechanical oscillators should not be very close
so that the dark modes can be broken efficiently. In addition, we use a second-order nonlinear medium (optical nonlinearity) to cool degenerate mechanical modes
even in the unresolved sideband regime. It is worth noting that there are two differences between our work and Ref. \cite{Huang20221}. First, in our scheme,
the dark modes of degenerate mechanical modes are broken by the Duffing nonlinearities of different mechanical modes. In Ref. \cite{Huang20221},
the dark modes are broken by an auxiliary cavity mode. Second, the ground-state cooling of all degenerate mechanical modes can be accomplished simultaneously even in
the unresolved sideband regime in the present work. However, we note that it is difficult to realize simultaneous ground-state cooling of all degenerate mechanical oscillators
in the unresolved sideband regime in Ref. \cite{Huang20221}.

The organization of this paper is as follows. In Sec. II, we introduce the model and
derive an effective Hamiltonian. In Sec. III, we derive the quantum Langevin equations of the present model.
In Sec. IV, we discuss the simultaneous ground-state cooling of two degenerate mechanical resonators in the unresolved sideband regime.
In Sec. V, we investigate the simultaneous ground-state cooling of three or four degenerate mechanical resonators.
In Sec. VI, we summarize our results.

\section{Model and Hamiltonian}

In the present work, we consider an optomechanical system formed by one optical cavity and $n$ mechanical resonators.
A second-order nonlinear medium $\chi^{(2)}$ is put into the Fabry-P\'{e}rot cavity. There are two modes in the optical cavity. One is the fundamental mode $a_1$ with frequency $\omega_c$.
The other is a second-order optical mode $a_2$ with frequencies $2\omega_c$. The decay rates of optical modes $a_1$ and $a_2$ are $\kappa_1$ and $\kappa_2$, respectively.
The mechanical oscillator $j$ with frequency $\omega_j$ and decay rate $\gamma_j$ is denoted by $b_j$. In addition, the fundamental and second-order modes are driven by two fields with amplitudes $\varepsilon_1$
and $\varepsilon_2$. The amplitude of the Duffing nonlinearity of mode $b_j$ is $\eta_j$. The schematic representation of our model can be found in Fig. 1.
The Hamiltonian of the system is ($\hbar = 1$) \cite{Lv20152,Asjad2019,Gan2019,Lau2020,Zhang20202}
\begin{eqnarray}
H &=& H_0 + H_{dr} + H_I + H_{\Lambda} + H_{\chi} ,\label{Hamiltonian1} \\
H_0 &=& \omega_c a_1^{\dag} a_1 + 2\omega_c a_2^{\dag} a_2 + \sum_{j = 1}^N\omega_j b_j^{\dag} b_j,\\
H_{dr} &=& i(\varepsilon_1 e^{-i\omega_L t} a_1^{\dag} + \varepsilon_2 e^{-2i\omega_L t} a_2^{\dag} - H.c.),\\
H_I &=& -\sum_{j=1}^N g_{1j} a_1^{\dag} a_1 (b_j^{\dag} + b_j) -\sum_{j=1}^N g_{2j} a_2^{\dag} a_2 (b_j^{\dag} + b_j),\\
H_{\Lambda} &=& \sum_{j=1}^N\frac{\eta_j}{2} (b_j^{\dag} + b_j)^4,\\
H_{\chi} &=& \frac{i \chi_0}{2} (a_1^{\dag 2} a_2 - a_1^2 a_2^{\dag}),
\end{eqnarray}
where $H_0$ is the free Hamiltonian of the present system and $H_{dr}$ is the Hamiltonian for driving fields
applied to the fundamental and second-order modes $a_1$ and $a_2$ with frequencies $\omega_L$ and $2\omega_L$.
$H_I$ denotes the interaction between the optical and mechanical modes with coupling constants $g_{ij}$.
$H_{\Lambda}$ is the Hamiltonian corresponding to the Duffing nonlinearities of mechanical modes. We note that a nonlinear amplitude of $\eta_j = 10^{-4}\omega_j$ ($j = 1,2,...,N$)
can be achieved by coupling the mechanical mode to an auxiliary system \cite{Lv20152}. The Hamiltonian of a second-order
nonlinear medium is denoted by $H_\chi$ and $\chi_0$ is the interaction
between the fundamental and second-order optical modes.
The amplitudes of driving fields are $\varepsilon_1 = \sqrt{2 \kappa_1 P_1 / \omega_L }$
and $\varepsilon_2 = \sqrt{2 \kappa_2 P_2 / (2\omega_L)}$ with $P_1$ and $P_2$ being the
powers of two driving lasers applied on optical modes $a_1$ and $a_2$.

In a rotating frame defined by the unitary transformation
$U(t) = \exp\{-i \omega_L t (a_1^{\dag} a_1+2 a_2^{\dag} a_2)\}$, the above Hamiltonian can be rewritten as
\begin{eqnarray}
H &=& U^{\dag} H U - i U^{\dag} \dot{U}\nonumber\\
 &=& \bar{\triangle}_c a_1^{\dag} a_1 + 2 \bar{\triangle}_c a_2^{\dag} a_2 + \sum_{j=1}^N\omega_j b_j^{\dag} b_j\nonumber\\
&& + i (\varepsilon_1 a_1^{\dag} - \varepsilon_1 a_1 + \varepsilon_2 a_2^{\dag} - \varepsilon_2 a_2)\nonumber\\
&& -\sum_{j=1}^N(g_{1j} a_1^{\dag} a_1 + g_{2j} a_2^{\dag} a_2) (b_j^{\dag} + b_j)\nonumber\\
&& + \sum_{j=1}^N\frac{\eta_j}{2} (b_j^{\dag} + b_j)^4 + \frac{i \chi_0}{2} (a_1^{\dag 2} a_2 - a_1^2 a_2^{\dag}),
\end{eqnarray}
with $\bar{\triangle}_c = \omega_c - \omega_L $.

\begin{figure}
\centering {\scalebox{0.4}[0.4]{\includegraphics{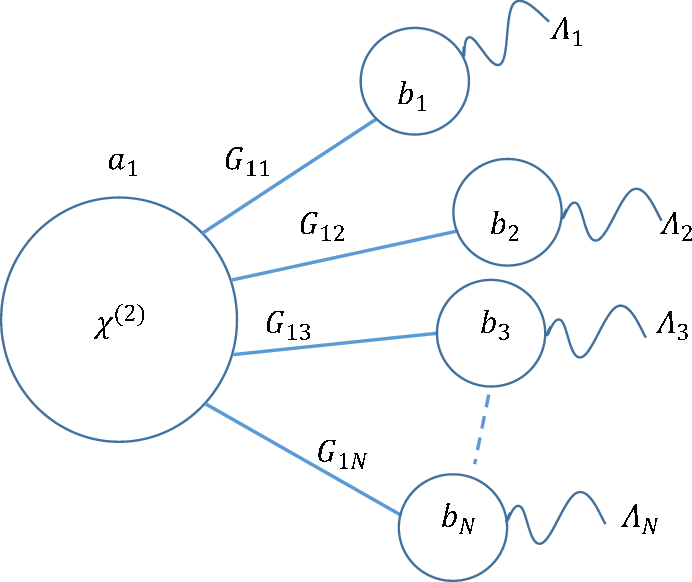}}}
\caption{Schematic representation of our model. The fundamental mode is represented by $a_1$ and mechanical mode $j$ is denoted by $b_j$. Here, the second-order optical mode $a_2$ is not shown.
A second-order nonlinear medium is denoted by $\chi^{(2)}$. The effective coupling strength between $a_1$ and $b_j$ is $G_{1j}$. $\Lambda_j$ is the Duffing nonlinearity (mechanical nonlinearity) of $b_j$. See Eqs. (\ref{Hamiltonian1}) and (\ref{QLEs}) for more details.
} \label{fig1}
\end{figure}

\section{Quantum Langevin equations}
In this section, we first linearize the above Hamiltonian by employing the following displacement transformations
$a_k \rightarrow \alpha_k + \delta a_k$ and $b_j \rightarrow \beta_j + \delta b_j$ with $k = 1,2$ and $j = 1,2,...,N$.
After some algebra, we obtain the quantum Langevin equations as follows
\begin{eqnarray}
\dot{\alpha_1} &=& -(i\Delta_c + \frac{\kappa_1}{2}) \alpha_1 + \chi_0 \alpha_1^\ast \alpha_2 + \varepsilon_1,\\
\dot{\alpha_2} &=& -(i \Delta_c' + \frac{\kappa_2}{2}) \alpha_2 - \frac{\chi_0}{2} \alpha_1^2 + \varepsilon_2,\\
\dot{\beta_j} &=& -(i \omega_j + \frac{\gamma_j}{2}) \beta_j + ig_{1j}|\alpha_1|^2 + ig_{2j}|\alpha_2|^2\nonumber\\
&& -i \eta_j(16|{\beta_j}|^3\cos^3{\varphi_{b_j}} + 12|{\beta_j}|\cos{\varphi_{b_j}}), \\
\delta \dot{a}_1 &=& -(i\Delta_c + \frac{\kappa_1}{2})\delta{a_1} + \chi_0\alpha_2 \delta a_1 ^\dagger + \chi_0\alpha_1^\ast\delta a_2 \nonumber\\
&& + i\sum_{j=1}^N G_{1j}(\delta b_j^{\dagger} + \delta b_j) + \sqrt{\kappa_1}a_{1,in},\\
\delta \dot{a}_2 &=& -(i\Delta_c' + \frac{\kappa_2}{2})\delta a_2  - \chi_0\alpha_1\delta a_1\nonumber\\
&& + i\sum_{j=1}^N G_{2j}(\delta b_j^\dagger + \delta b_j) + \sqrt{\kappa_2}a_{2,in},\\
\delta \dot{b}_j &=& -(i\omega_j + \frac{\gamma_j}{2})\delta b_j + i(G_{1j}\delta a_1^\dagger + G_{1j}^\ast\delta a_1) \nonumber\\
&& + i(G_{2j}\delta a_2^\dagger + G_{2j}^\ast\delta a_2)\nonumber\\
&& -2i\Lambda_j(\delta b_j^\dagger + \delta b_j) + \sqrt{\gamma_j}b_{j,in},
\end{eqnarray}
where $\Delta_c = \bar{\Delta}_c-2\sum_{j=1}^N g_{1j}|\beta_j|\cos{\varphi_{b_j}}$, $\Delta_c' = 2\bar{\Delta}_c-2\sum_{j=1}^N g_{2j}|\beta_j|\cos{\varphi_{b_j}}$,
$\Lambda_j = 3\eta_j(4|\beta_j|^2\cos^2{\varphi_{b_j}} + 1)$, and $G_{kj} = g_{kj}\alpha_{k}$ ($k=1,2$ and $j = 1,2,...,N$). We have assumed $\beta_{j} = |\beta_{j}|e^{i\varphi_{b_{j}}}$.

Note that the fluctuations of mode $a_2$ can be neglected in the limit of large $\kappa_2$ and the adiabatic
approximation is valid \cite{Asjad2019,Zhang20202}. Thus the quantum Langevin equations can be rewritten as
\begin{eqnarray}
\delta \dot{a}_1 &=& -(i\Delta_c + \frac{\kappa_1}{2})\delta a_1 + i\sum_{j=1}^N G_{1j}(\delta b_j^\dagger + \delta b_j)\nonumber\\
&& + \chi \delta a_1^\dagger + \sqrt{\kappa_1}a_{1,in},\nonumber\\
\delta \dot{b}_j &=& -(i\omega_j + \frac{\gamma_j}{2})\delta b_j + i G_{1j}(\delta a_1^\dagger + \delta a_1)\nonumber\\
&& -2i\Lambda_j(\delta b_j^\dagger + \delta b_j) + \sqrt{\gamma_j}b_{j,in}, \label{QLEs}
\end{eqnarray}
where $\chi = \chi_0\alpha_2 = |\chi|e^{2i\varphi}$. Without loss of generality, $G_{kj}$ has been assumed to be real.

We define the quadrature operators $X_{O = a_1,b_j} = (\delta O^\dagger + \delta O)/\sqrt{2}$, $Y_{O = a_1,b_j} = i(\delta O^\dagger - \delta O)/\sqrt{2}$,
and the noise quadrature operators $X_{O = a_1,b_j}^{in} = (O_{in}^\dagger + O_{in})/\sqrt{2}$ and $Y_{O = a_1,b_j}^{in} = i(O_{in}^\dagger - O_{in})/\sqrt{2}$.
From the above quantum Langevin equations, we obtain
\begin{eqnarray}
\dot{\vec{f}} = A\vec{f} + \vec{n}, \label{dfdt}
\end{eqnarray}
where $\vec{f} = (X_{a_1},Y_{a_1},X_{b_1},Y_{b_1},..., X_{b_N},Y_{b_N})^T$ and
\begin{eqnarray}
\vec{n} &=& (\sqrt{\kappa_1}X_{a_1}^{in},\sqrt{\kappa_1}Y_{a_1}^{in},\sqrt{\gamma_1}X_{b_1}^{in}, \sqrt{\gamma_1}Y_{b_1}^{in},..., \nonumber\\
&& \sqrt{\gamma_N}X_{b_N}^{in},\sqrt{\gamma_N}Y_{b_N}^{in})^T,
\end{eqnarray}

\begin{widetext}
\begin{eqnarray}
A =
\left(
\begin{array}{ccccccc}
|\chi|\cos{(2\phi)} - \frac{\kappa_1}{2}  & |\chi|\sin{(2\phi)} + \Delta_c            &         0            &            0        & ... &        0           & 0 \\
|\chi|\sin{(2\phi)} - \Delta_c            & -|\chi|\cos{(2\phi)} - \frac{\kappa_1}{2} &       2G_{11}        &            0        & ... &      2G_{1N}       & 0 \\
             0                            &                     0                     &  -\frac{\gamma_1}{2} &        \omega_1     & ... &        0           & 0 \\
             2G_{11}                      &                     0                     & -\omega_1-4\Lambda_1 & -\frac{\gamma_1}{2} & ... &        0           & 0 \\
             ...                          &                    ...                    &        ...           &           ...       & ... &       ...          & ...\\
             0                            &                     0                     &         0            &            0        & ... & -\frac{\gamma_N}{2} & \omega_N \\
             2G_{1N}                      &                     0                     &         0            &            0        & ... & -\omega_N-4\Lambda_N & -\frac{\gamma_N}{2} \\
\end{array}
\right).
\end{eqnarray}
\end{widetext}
The dynamics of the system described by Eq.(\ref{dfdt}) can be completely described by a $2 (N + 1)\times 2(N + 1)$ covariance matrix $V$ with $V_{j, k} = \langle f_jf_k + f_kf_j \rangle/2$.
Here, $N$ is the number of mechanical modes.
We obtain the evolution of the covariance matrix $V$ as follows:
\begin{eqnarray}
\dot{V} = AV + VA^T + D, \label{dVdt}
\end{eqnarray}
where $D$ is the noise correlation defined by
$D = diag[\frac{\kappa_1}{2},\frac{\kappa_1}{2},\frac{\gamma_1}{2}(2n_{th} + 1),\frac{\gamma_1}{2}(2n_{th} + 1),...,\frac{\gamma_N}{2}(2n_{th} + 1),\frac{\gamma_N}{2}(2n_{th} + 1)]$. Here $n_{th}$ is the mean phonon number of the mechanical resonators.
According to the Routh-Hurwitz criterion \cite{Dejesus1987}, the system described by Eq.(\ref{dVdt}) is stable only if all the real parts of the
eigenvalues of the matrix A are negative. All the parameters used in the present work satisfy the Routh-Hurwitz criterion.
The steady-state mean phonon numbers of mechanical modes $b_j$ are
\begin{eqnarray}
n_j &=& (V_{2j+1, 2j+1} + V_{2j+2, 2j+2} - 1)/2.
\end{eqnarray}

Now, we discuss the strengths of the mechanical and optical nonlinearities. In general, the natural Duffing nonlinearity (mechanical nonlinearity) is very small.
For example, in Ref. \cite{Lv20152}, the authors have shown that a nonlinear amplitude of $\eta_j = 10^{-4}\omega_j$
can be achieved by coupling the mechanical mode to an auxiliary system. Here, $\omega_j$ is the frequency of the mechanical mode $b_j$.
Also, the optical nonlinearity strength is, in general, small compared with the frequency of the mechanical mode $\omega_j$.
Note that the effective mechanical nonlinearity $\Lambda_j$ and the effective optical nonlinearity $\chi$ in Eqs. (13) and (14)
are defined by $\Lambda_j = 3\eta_j(4|\beta_j|^2\cos^2{\varphi_{b_j}} + 1)$ and $\chi = \chi_0\alpha_2$, respectively.
It is clear to see the effective nonlinearities $\Lambda_j$ and $\chi$ could be significantly enhanced by $\beta_j$ and $\alpha_2$.
In Fig. 2, we plot the steady-state amplitudes $|\alpha_2|$ and $|\beta_j|$ as functions of the driving power $P$ using Eqs. (8)-(10). We assume $P_1 = P_2=P$,
$\omega_1 = \omega_2 =...=\omega_N$, and $\gamma_1 = \gamma_2 = ...=\gamma_N$ in this figure.
From Fig.2, one can see that $|\alpha_2| \approx 400 $ and $|\beta_j| \approx 13.7$ if $P = 4 \mu W$.
The natural Duffing nonlinearity strength is $\eta_j = 10^{-4}\omega_1$ and the optical nonlinearity strength is $\chi_0 = 10^{-3}\omega_1$.
In the case of $P = 4\mu W$ and $\varphi_{b_j} = 0$, we find $\Lambda_j = 3\eta_j(4|\beta_j|^2\cos^2{\varphi_{b_j}} + 1) \approx 0.23 \omega_1$ and $|\chi| = \chi_0|\alpha_2| \approx 0.4\omega_1$.
In a word, the effective nonlinearities $\Lambda_j$ and $\chi$ could be significantly enhanced in the present model.

\begin{figure}[tbp]
\centering {\scalebox{0.4}[0.4]{\includegraphics{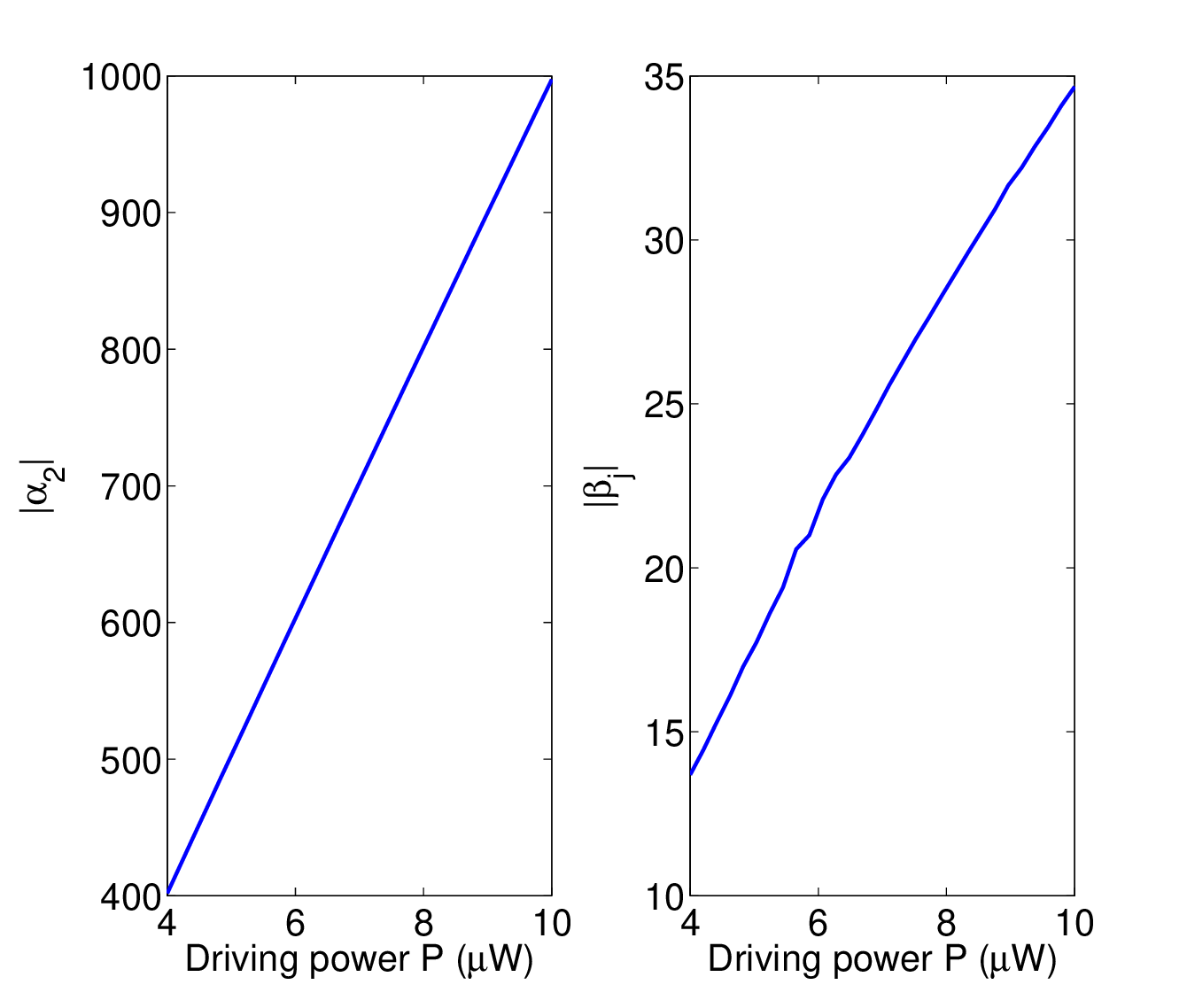}}}
\caption{ Steady-state amplitudes $|\alpha_2|$ and $|\beta_j|$ are plotted as functions
of the driving power $P$ with $\eta = 10^{-4}\omega_1$ and $|\chi_0| = 10^{-3}\omega_1$.
We assume $P_1 = P_2= P$ in the present work. Other parameter
values are $\omega_1 = \omega_2 = ... =\omega_N = 2\pi \times 20MHz$, $\omega_c = 2\pi \times 500THz$, $G_{1j} = G_{2j} = 10^{-4}\omega_1$ ($j = 1,2,...,N$),
$\kappa_1 = 100\omega_1$, $\kappa_2 = 2000\omega_1$, $\gamma_1 = \gamma_2 = ... =\gamma_N = 10^{-6} \omega_1$, $\varphi_{b_j} = 0$, $\Delta_c = 10\omega_1$, and $\Delta_c' = 20\omega_1$.
} \label{fig2}
\end{figure}

\section{Simultaneous ground-state cooling of two degenerate mechanical oscillators}
In this section, we investigate the simultaneous ground-state cooling of two degenerate mechanical resonators beyond the resolved sideband regime with
the help of optical and mechanical nonlinearities. First, we derive the linearized Hamiltonian under the rotating wave approximation (RWA) in the absence of
mechanical nonlinearities. Second, we recast the linearized Hamiltonian based on two hybrid mechanical modes $B_1$ and $B_2$. The dark mode effect can be seen clearly from this
Hamiltonian in the absence of mechanical nonlinearities. Third, we introduce the mechanical nonlinearities. One can find that the dark mode formed by two degenerate mechanical resonators
is destroyed by the mechanical nonlinearities with different amplitudes. Finally, the optical nonlinearity is used to cool degenerate mechanical resonators simultaneously
even in the unresolved sideband regime.

\subsection{Dark mode of two degenerate mechanical oscillators}
From Eqs. (\ref{QLEs}), we can write out the linearized Hamiltonian under the RWA explicitly as
\begin{eqnarray}
\tilde{H}_{RWA} &=& \tilde{H}_0 + \tilde{H}_I + \tilde{H}_\Lambda + \tilde{H}_\chi,\nonumber\\
\tilde{H}_0 &=& \Delta_c\delta a_1^\dagger \delta a_1 + \omega_1\delta b_1^\dagger \delta b_1 + \omega_2\delta b_2^\dagger \delta b_2,\nonumber\\
\tilde{H}_I &=& - G_{11}(\delta a_1^\dagger \delta b_1 + \delta b_1^\dagger \delta a_1) \nonumber\\
            &&- G_{12}(\delta a_1^\dagger\delta b_2 + \delta b_2^\dagger \delta a_1),\nonumber\\
\tilde{H}_\Lambda &=& \Lambda_1(\delta b_1^{\dagger 2} + \delta b_1^2 + 2\delta b_1^\dagger \delta b_1) \nonumber\\
                  &&+ \Lambda_2(\delta b_2^{\dagger 2} + \delta b_2^2 + 2\delta b_2^\dagger \delta b_2),\nonumber\\
\tilde{H}_\chi &=& \frac{i\chi}{2}(\delta a_1^{\dagger 2} - \delta a_1^2).
\end{eqnarray}
We now introduce two hybrid mechanical modes $B_1$ and $B_2$ as follows:
\begin{eqnarray}
B_1 &=& \frac{1}{\sqrt{{G_{11}^2} + G_{12}^2}}(G_{11}\delta{b_1} + G_{12}\delta{b_2}),\nonumber\\
B_2 &=& \frac{1}{\sqrt{{G_{11}^2} + G_{12}^2}}(G_{12}\delta{b_1} - G_{11}\delta{b_2}).
\end{eqnarray}
Here, the new operators $B_1$ and $B_2$ satisfy the bosonic commutation relations $[B_1,{B_1}^\dagger] = 1$ and $[B_2,{B_2}^\dagger] = 1.$
The Hamiltonian $\tilde{H}_{RWA}$ can be expressed as
\begin{eqnarray}
\tilde{H}_{RWA} &=& \tilde{H'}_0 + \tilde{H'}_I + \tilde{H'}_{\Lambda} + \tilde{H'}_{\chi},\nonumber\\
\tilde{H'}_0 &=& \Delta_c\delta a_1^\dagger\delta a_1 + \omega_{1,w} B_1^\dagger B_1 + \omega_{2,w} B_2^\dagger B_2,\nonumber\\
\tilde{H'}_I &=& G_+(\delta a_1^\dagger B_1 + \delta a_1{B_1}^\dagger) + \xi_w({B_1}^\dagger B_2 + {B_2}^\dagger B_2),\nonumber\\
\tilde{H'}_\Lambda &=& \omega_{1,\Lambda}(B_1^2 + {B_1^\dagger}^2 + 2B_1^\dagger B_1) \nonumber\\
&&+ \omega_{2,\Lambda}(B_2^2 + B_2^{\dagger 2} + 2B_2^\dagger B_2)\nonumber\\
&&+ \xi_{\Lambda}(B_1B_2 + B_2B_1 + B_1^\dagger B_2^\dagger + B_2^\dagger B_1^\dagger \nonumber\\
&&+ 2B_1^\dagger B_2 + 2B_2^\dagger B_1),\nonumber\\
\tilde{H'}_\chi &=&\frac{i\chi}{2}(\delta a_1^{\dagger 2} - \delta a_1^2), \label{H_RWA}
\end{eqnarray}
with
\begin{eqnarray}
\omega_{1,w} &=& \frac{\omega_1G_{11}^2 + \omega_2G_{12}^2}{G_{11}^2 + G_{12}^2},\nonumber\\
\omega_{2,w} &=& \frac{\omega_1G_{12}^2 + \omega_2G_{11}^2}{G_{11}^2 + G_{12}^2},   \nonumber\\
\omega_{1,\Lambda} &=& \frac{\Lambda_1G_{11}^2 + \Lambda_2G_{12}^2}{G_{11}^2 + G_{12}^2},\nonumber\\
\omega_{2,\Lambda} &=& \frac{\Lambda_1G_{12}^2 + \Lambda_2G_{11}^2}{G_{11}^2 + G_{12}^2},\nonumber\\
G_+ &=& \sqrt{G_{11}^2 + G_{12}^2},\nonumber\\
\xi_w &=& \frac{(\omega_1 - \omega_2)G_{11}G_{12}}{G_{11}^2 + G_{12}^2},  \nonumber\\
\xi_{\Lambda} &=& \frac{(\Lambda_1 - \Lambda_2)G_{11}G_{12}}{G_{11}^2 + G_{12}^2}. \label{xi}
\end{eqnarray}

Some remarks must be made now.
First, if the frequencies of two mechanical modes are equal ($\omega_1 = \omega_2$), then, one can easily find $\xi_{w} = 0$.
Second, in the absence of mechanical nonlinearities ($\Lambda_1 = \Lambda_2 = 0$) or
the mechanical nonlinearities are equal ($\Lambda_1 = \Lambda_2 > 0$), the term $\xi_{\Lambda} = 0$.
Thus, the hybrid mode $B_2$ is totally decoupled from $B_1$ and $\delta a_1$ simultaneously in the case of $\omega_1 = \omega_2$ ($\xi_w = 0$) and $\Lambda_1 = \Lambda_2$ ($\xi_{\Lambda} = 0$) as we can see from Eqs. (\ref{H_RWA}) and (\ref{xi}).
The hybrid mode $B_2$ is called a dark mode \cite{Huang20221,Huang20222} and cannot be cooled completely.
In other words, the mechanical system cannot be cooled if the frequencies of two mechanical resonators are equal ($\xi_w = 0$) and the mechanical nonlinearities are also equal ($\xi_{\Lambda} = 0$).

The reason is as follows.
The thermal excitations of $B_1$ can be extracted through the cooling channel of the optical modes $\delta a_1$ since there is direct interaction between $B_1$ and $\delta a_1$ as one can clearly see from the term $G_+(\delta a_1^\dagger B_1 + \delta a_1{B_1}^\dagger)$ of Eq. (\ref{H_RWA}).
Unfortunately, the thermal energy of the dark mode $B_2$ cannot be extracted through the cooling channel if $\xi_w = 0$ and $\xi_{\Lambda} = 0$, and there is no direct interaction between $B_2$ and
$B_1$ or between $B_2$ and $\delta a_1$.
Physically, the dark mode formed by two degenerate mechanical modes can be destroyed by introducing mechanical nonlinearities with different amplitudes, i.e., $\Lambda_1 \neq \Lambda_2 $ ($\xi_{\Lambda} \neq 0$). In this case, $\xi_w = 0$, while $\xi_{\Lambda} \neq 0$. Therefore, there is direct interaction between $B_2$ and $B_1$. The thermal excitations of $B_1$ and $B_2$ can be efficiently extracted
through the cooling channel of the optical mode $\delta a_1$. It is possible to realize
simultaneous ground-state cooling of degenerate mechanical modes with the help of mechanical nonlinearities when the amplitudes are not very close.

\subsection{Cooling beyond resolved sideband regime}
In general, in the standard sideband cooling, one obstacle for ground-state cooling of mechanical resonators is the resolved sideband condition, i.e.,
the decay rate of an optical cavity must be smaller than the frequencies of mechanical modes \cite{Bowen2015}.
This restriction typically requests that the finesse of an optical cavity should be very high, which limits
the size of mechanical resonators to be cooled. In the Hamiltonian $\tilde{H'}_\chi$ of Eqs. (\ref{H_RWA}), the optical nonlinearity $\chi$ is employed to cooled mechanical resonators beyond the
resolved sideband regime.

\begin{figure}[tbp]
\centering {\scalebox{0.4}[0.4]{\includegraphics{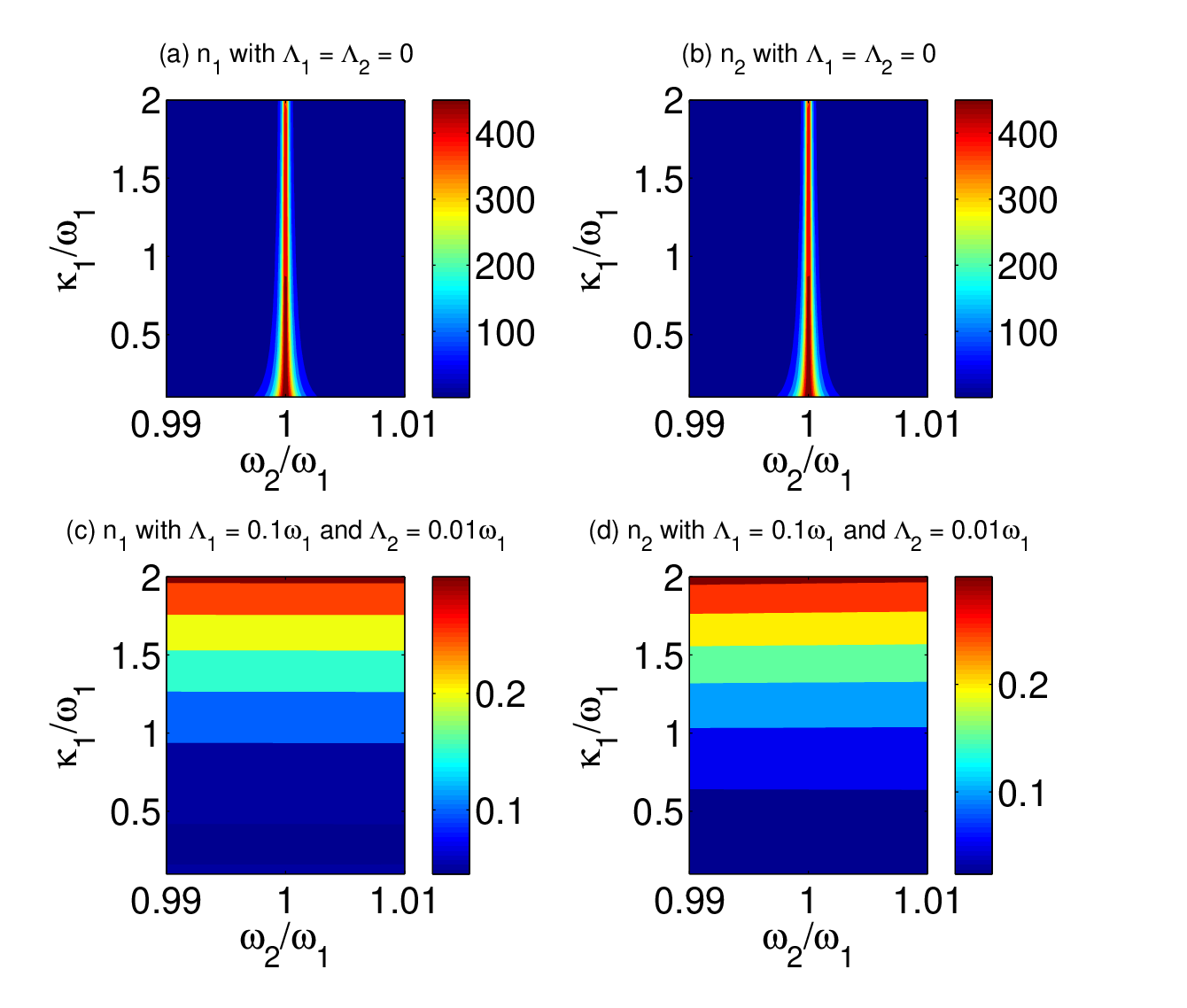}}}
\caption{Steady-state mean phonon numbers $n_1$ and $n_2$ versus $\omega_2/\omega_1$ and $\kappa_1/\omega_1$. The parameters are $\gamma_1 = \gamma_2 = 10^{-6} \omega_1$, $G_{11} = G_{12} = 0.1\omega_1, \phi = 0.5\pi,
\Delta_c = \omega_1$, $|\chi| = 0$, and $n_{th} = 1000$.
} \label{fig3}
\end{figure}

\begin{figure}[tbp]
\centering {\scalebox{0.4}[0.4]{\includegraphics{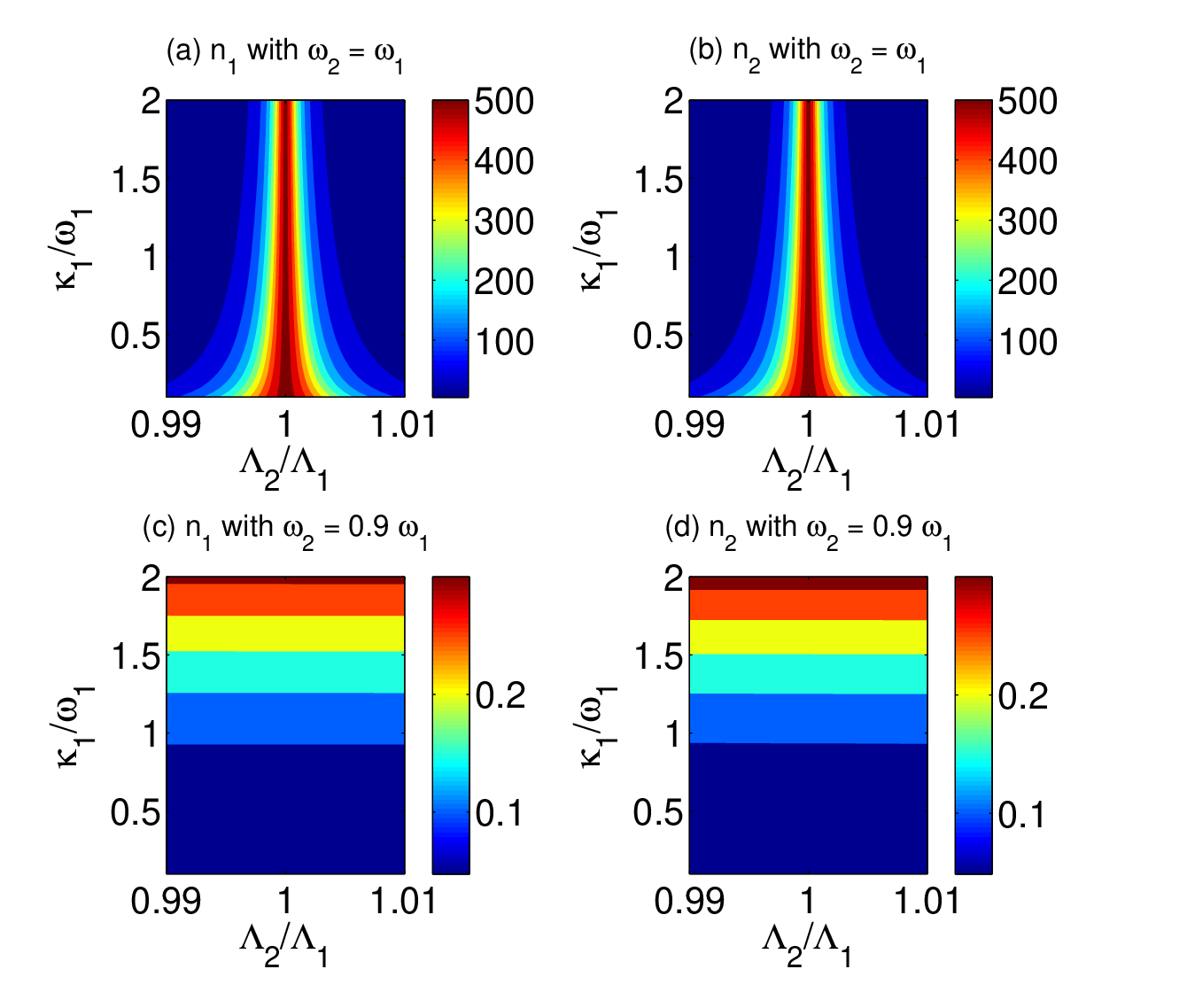}}}
\caption{Steady-state mean phonon numbers $n_1$ and $n_2$ versus $\Lambda_2/\Lambda_1$ and $\kappa_1/\omega_1$. The parameters are $\gamma_1 = \gamma_2 = 10^{-6} \omega_1$, $G_{11} = G_{12} = 0.1\omega_1, \phi = 0.5\pi,
\Delta_c = \omega_1$, $|\chi| = 0$, $\Lambda_1 = 0.1\omega_1$, and $n_{th} = 1000$.
} \label{fig4}
\end{figure}

In Fig. 3, we plot the steady-state mean phonon numbers of mechanical modes $b_1$ and $b_2$ as functions of parameters $\omega_2/\omega_1$ and $\kappa_1/\omega_1$ for different $\Lambda_1$ and $\Lambda_2$.
From Fig. 3(a) and Fig. 3(b), one can see that, in the absence of mechanical nonlinearities, the mean phonon numbers of the first and second mechanical modes
can be much larger than 1 if the frequency of the second mechanical mode is close to the frequency of the first mechanical mode, i.e., $\omega_2 \approx \omega_1$. Clearly, this is a natural consequence of the dark-mode effect, i.e., when the frequencies of two mechanical modes are equal, they form a hybrid dark mode $B_2$ as we have pointed out previously. If the mechanical nonlinearities with different amplitudes are introduced, the steady-state mean phonon numbers of two degenerate mechanical resonators can be cooled efficiently as one can find in Fig. 3(c) and Fig. 3(d). This implies that the mechanical nonlinearities with different amplitudes can break the dark mode $B_2$ since the parameter $\xi_{\Lambda}$ in Eq. (\ref{H_RWA}) is not zero and there is interaction between hybrid modes $B_1$ and $B_2$. The thermal excitations of $B_2$ can be extracted via the cooling channel $B_2 \Rightarrow B_1 \Rightarrow a_1$ as one can clearly see from Eq. (\ref{H_RWA}).

In Fig. 4, we plot the stationary-state mean phonon numbers $n_1$ and $n_2$ as functions of parameters $\Lambda_2/\Lambda_1$ and $\kappa_1/\omega_1$ for different value of $\omega_2$.
From Fig. 4(a) and Fig. 4(b), we find that two degenerate mechanical resonators $b_1$ and $b_2$ can not be
cooled if the amplitudes of two mechanical nonlinearities are close.
From Fig. 4(c) and Fig. 4(d), one can see the two mechanical resonators can be cooled simultaneously if the
difference between their frequencies is not very small. This is consistent with the results of \cite{Genes2008}\cite{Ockeloen2019}.

\begin{figure}[tbp]
\centering {\scalebox{0.4}[0.4]{\includegraphics{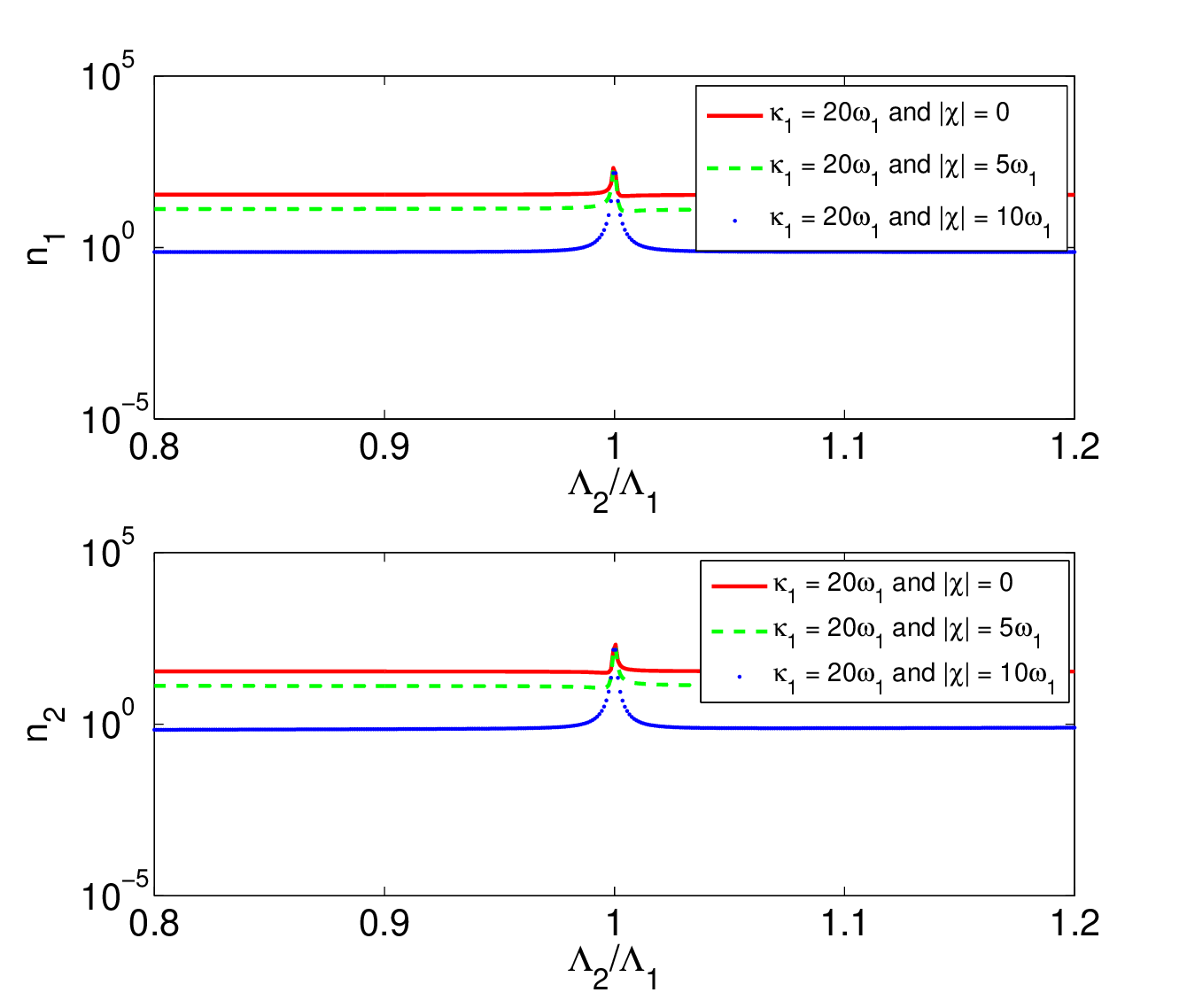}}}
\caption{Steady-state mean phonon numbers $n_1$ and $n_2$ versus $\Lambda_2/\Lambda_1$ for different $\kappa_1$ and $|\chi|$. The parameters are $\gamma_1 = \gamma_2 = 10^{-6} \omega_1$, $G_{11} = G_{12} = 0.1\omega_1, \phi = 0.5\pi,
\Delta_c = \omega_1$, $\Lambda_1 = 0.1\omega_1$, $\omega_2 = \omega_1$, and $n_{th} = 1000$.
} \label{fig5}
\end{figure}

Now, we turn to investigate the simultaneous ground-state cooling of two degenerate mechanical resonators
in the highly unresolved sideband regime with the help of optical nonlinearity.
In Fig. 5, we plot the mean phonon numbers as functions of $\Lambda_2/\Lambda_1$ for different values of $\kappa_1$ and $|\chi|$. In the absence of optical nonlinearity, the mean phonon numbers $n_1$ and $n_2$
are always lager than 10 when the decay rate of the optical mode is much larger than $\omega_1$ [see the red solid lines of Fig. 5(a) and Fig. 5(b)]. If the optical nonlinearity is introduced, then, the mean phonon numbers are
reduced [see the green dashed lines of Fig. 5(a) and Fig. 5(b)]. However, $n_1$ and $n_2$ are larger than 10 in the case of $|\chi| = 5\omega_1$ and $\kappa_1 = 20\omega_1$. Particularly, $n_1$ and $n_2$ can be smaller than 1 in the case of $|\chi| = 10\omega_1$
and the difference of two mechanical nonlinearities is not very close [see the blue dotted lines of Fig. 5(a) and Fig. 5(b)].

\begin{figure}[tbp]
\centering {\scalebox{0.4}[0.4]{\includegraphics{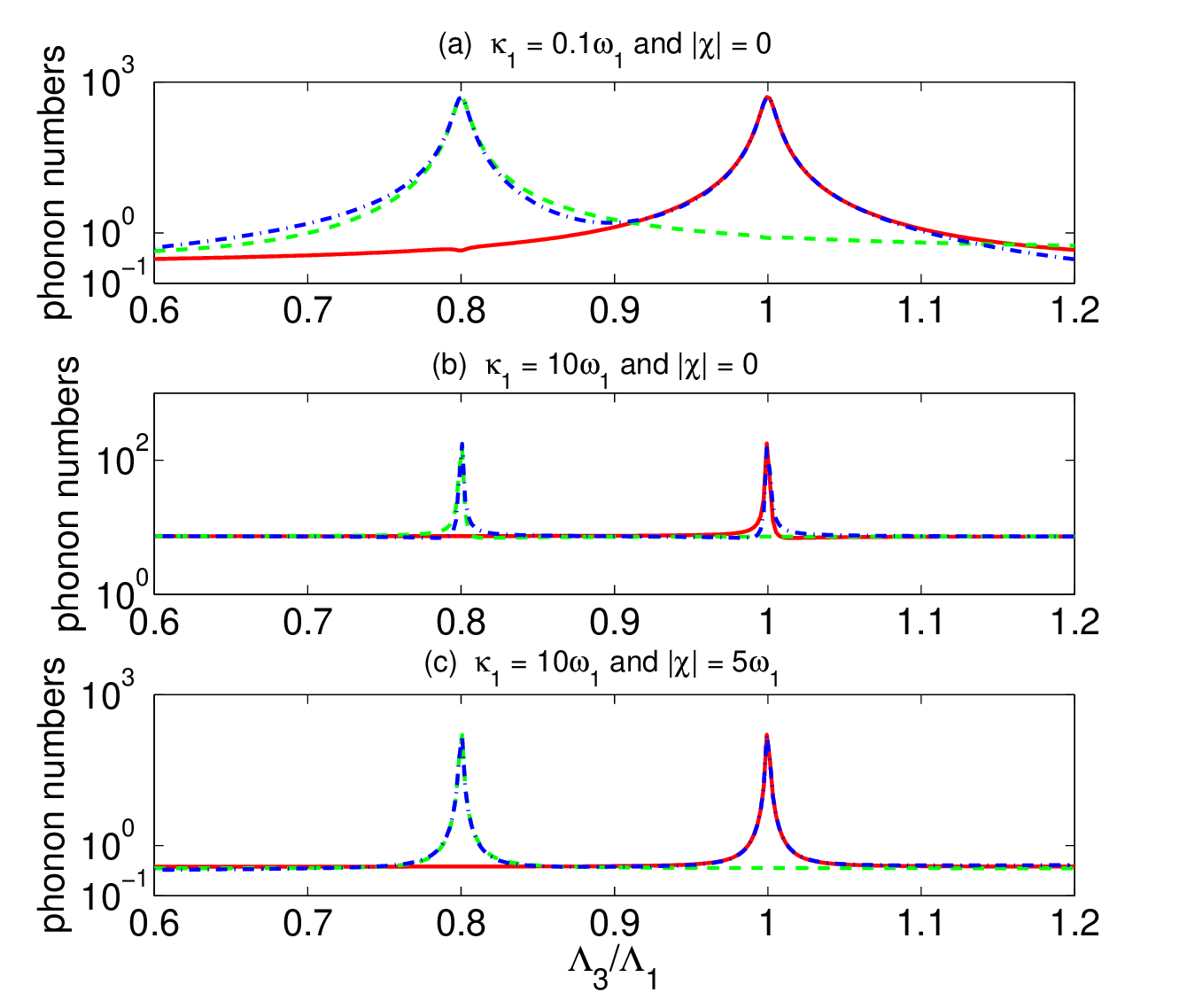}}}
\caption{Steady-state mean phonon numbers of mechanical modes $n_1$ (red solid lines),
$n_2$ (green dashed lines), and $n_3$ (blue dash-dot lines) versus $\Lambda_3/\Lambda_1$ for different $\kappa_1$ and $|\chi|$. The parameters are $\gamma_1 = \gamma_2 = \gamma_3 = 10^{-6} \omega_1$, $G_{11} = G_{12} = G_{13} = 0.3\omega_1, \phi = 0.5\pi,
\Delta_c = \omega_1$, $\Lambda_1 = 0.1\omega_1$, $\Lambda_2 = 0.8\Lambda_1$, $\omega_1 = \omega_2 = \omega_3$, and $n_{th} = 1000$.
} \label{fig6}
\end{figure}

\section{Simultaneous ground-state cooling of three or four degenerate mechanical oscillators}
In this section, we study the simultaneous ground-state cooling of three or four degenerate mechanical modes in the
unresolved sideband regime using mechanical nonlinearities with different amplitudes and optical nonlinearity.
\subsection{Three degenerate mechanical oscillators}

Similar to the previous section, we obtain the quantum Langevin equations for the fundamental optical mode $a_1$ and three mechanical modes $b_1$, $b_2$, and $b_3$ as
\begin{eqnarray}
\delta \dot{a}_1 &=& -(i\Delta_c + \frac{\kappa_1}{2})\delta a_1 + iG_{11}(\delta b_1^\dagger + \delta b_1) \nonumber\\
&& + iG_{12}(\delta b_2^\dagger + \delta b_2) + iG_{13}(\delta b_3^\dagger + \delta b_3)\nonumber\\
&& + \chi \delta a_1^\dagger + \sqrt{\kappa_1}a_{1,in},\nonumber\\
\delta \dot{b}_j &=& -(i\omega_j + \frac{\gamma_j}{2})\delta b_j + i G_{1j}(\delta a_1^\dagger + \delta a_1)\nonumber\\
&& -2i\Lambda_j(\delta b_j^\dagger + \delta b_j) + \sqrt{\gamma_j}b_{j,in}, \label{QLEs_3}
\end{eqnarray}
with $j=1,2,3$. The quantum Langevin equations are still given by Eq.(\ref{dfdt}) with
$\vec{f} = (X_{a_1},Y_{a_1},X_{b_1},Y_{b_1},X_{b_2},Y_{b_2},X_{b_3},Y_{b_3})^T$ and
\begin{eqnarray}
\vec{n} &=& (\sqrt{\kappa_1}X_{a_1}^{in},\sqrt{\kappa_1}Y_{a_1}^{in},\sqrt{\gamma_1}X_{b_1}^{in}, \sqrt{\gamma_1}Y_{b_1}^{in}, \nonumber\\
&& \sqrt{\gamma_2}X_{b_2}^{in},\sqrt{\gamma_2}Y_{b_2}^{in},\sqrt{\gamma_3}X_{b_3}^{in},\sqrt{\gamma_3}Y_{b_3}^{in})^T.
\end{eqnarray}

Similar to the previous section, the steady-state mean phonon numbers of mechanical modes $b_1$, $b_2$, and $b_3$ are
\begin{eqnarray}
n_1 &=& (V_{33} + V_{44} - 1)/2, \nonumber\\
n_2 &=& (V_{55} + V_{66} - 1)/2, \nonumber\\
n_3 &=& (V_{77} + V_{88} - 1)/2,
\end{eqnarray}
and $V$ is a $8\times8$ covariance matrix with $V_{jk} = \langle f_jf_k + f_kf_j \rangle/2$.
The dynamics of the covariance matrix $V$ is similar to Eq.(\ref{dVdt}) with
\begin{widetext}
\begin{eqnarray}
A &=&
\left(
\begin{array}{cccccccc}
|\chi|\cos{(2\phi)} - \frac{\kappa_1}{2}  & |\chi|\sin{(2\phi)} + \Delta_c            &         0            &            0        &          0           & 0          &   0               &   0\\
|\chi|\sin{(2\phi)} - \Delta_c            & -|\chi|\cos{(2\phi)} - \frac{\kappa_1}{2} &       2G_{11}        &            0        &        2G_{12}       & 0          & 2G_{13}           &   0\\
             0                            &                     0                     &  -\frac{\gamma_1}{2} &        \omega_1     &          0           & 0          &0                  &   0 \\
             2G_{11}                      &                     0                     & -\omega_1-4\Lambda_1 & -\frac{\gamma_1}{2} &          0           & 0          &0                  &   0 \\
             0                            &                     0                     &         0            &            0        &  -\frac{\gamma_2}{2} & \omega_2   &0                  &   0 \\
             2G_{12}                      &                     0                     &         0            &            0        & -\omega_2-4\Lambda_2 & -\frac{\gamma_2}{2} & 0        &   0 \\
             0                            &                     0                     &         0            &            0        &          0           & 0  & -\frac{\gamma_3}{2}       & \omega_3\\
             2G_{13}                      &                     0                     &         0            &            0        &          0           & 0  &-\omega_3-4\Lambda_3       & -\frac{\gamma_3}{2}\\
\end{array}
\right), \\
D &=& diag[\frac{\kappa_1}{2},\frac{\kappa_1}{2},\frac{\gamma_1}{2}(2n_{th} + 1),\frac{\gamma_1}{2}(2n_{th} + 1),\frac{\gamma_2}{2}(2n_{th} + 1),\frac{\gamma_2}{2}(2n_{th} + 1),\frac{\gamma_3}{2}(2n_{th} + 1),\frac{\gamma_3}{2}(2n_{th} + 1)].
\end{eqnarray}
\end{widetext}

In Fig. 6, we plot steady-state mean phonon numbers $n_1$, $n_2$, and $n_3$ as functions of $\Lambda_3/\Lambda_1$
with $\Lambda_2 = 0.8\Lambda_1$ and $\omega_1 = \omega_2 = \omega_3$. In other words, the frequencies of
three mechanical modes are the same. However, the mechanical nonlinearities of $b_1$ and $b_2$ are not very close
since $\Lambda_2 = 0.8\Lambda_1$. As a result, the mechanical modes $b_1$ and $b_2$ cannot form a dark mode
since the dark mode can be destroyed by two mechanical nonlinearities with different amplitudes.
Note that $\Lambda_3$ could be equal to $\Lambda_2$ or $\Lambda_1$ and it is still possible to form a dark mode
between $b_1$ and $b_3$ ($\Lambda_1 = \Lambda_3$) or $b_2$ and $b_3$ ($\Lambda_2 = \Lambda_3$).
In Fig. 6(a), we calculate the mean phonon numbers in the resolved sideband regime with $\kappa_1 = 0.1\omega_1$
and the optical nonlinearity is assumed to be zero with $|\chi| = 0$.
There are two peaks of $n_3$ (blue dash-dot lines). One peak is located at $\Lambda_3 = \Lambda_2 = 0.8\Lambda_1$,
which is a result of the dark mode formed by $b_2$ and $b_3$. The first mechanical mode $b_1$ can be cooled efficiently
even in the presence of the dark mode which is mixed by $b_2$ and $b_3$ in the resolved sideband regime.
The other peak is located at $\Lambda_3 = \Lambda_1$, which is a result of the dark mode formed by $b_1$ and $b_3$.
In this case, $b_1$ and $b_3$ cannot be cooled, while $n_2$ could be smaller than 1.
In a word, if we want to achieve ground-state cooling of all mechanical modes, the mechanical nonlinearities
should not be too close. If the optical decay rate is much larger than the frequencies of mechanical modes,
all the stationary-state mean phonon numbers of mechanical modes are larger than 1 when $\kappa_1 = 10\omega_1$
and $|\chi| = 0$ as one can clearly see from Fig. 6(b). Therefore, we introduce the optical nonlinearity.
In Fig. 6(c), we find all the degenerate mechanical resonators can be cooled efficiently even in the
unresolved sideband regime with the help of optical nonlinearity with $|\chi| = 5\omega_1$.

\subsection{Four degenerate mechanical oscillators}

\begin{figure}[tbp]
\centering {\scalebox{0.4}[0.4]{\includegraphics{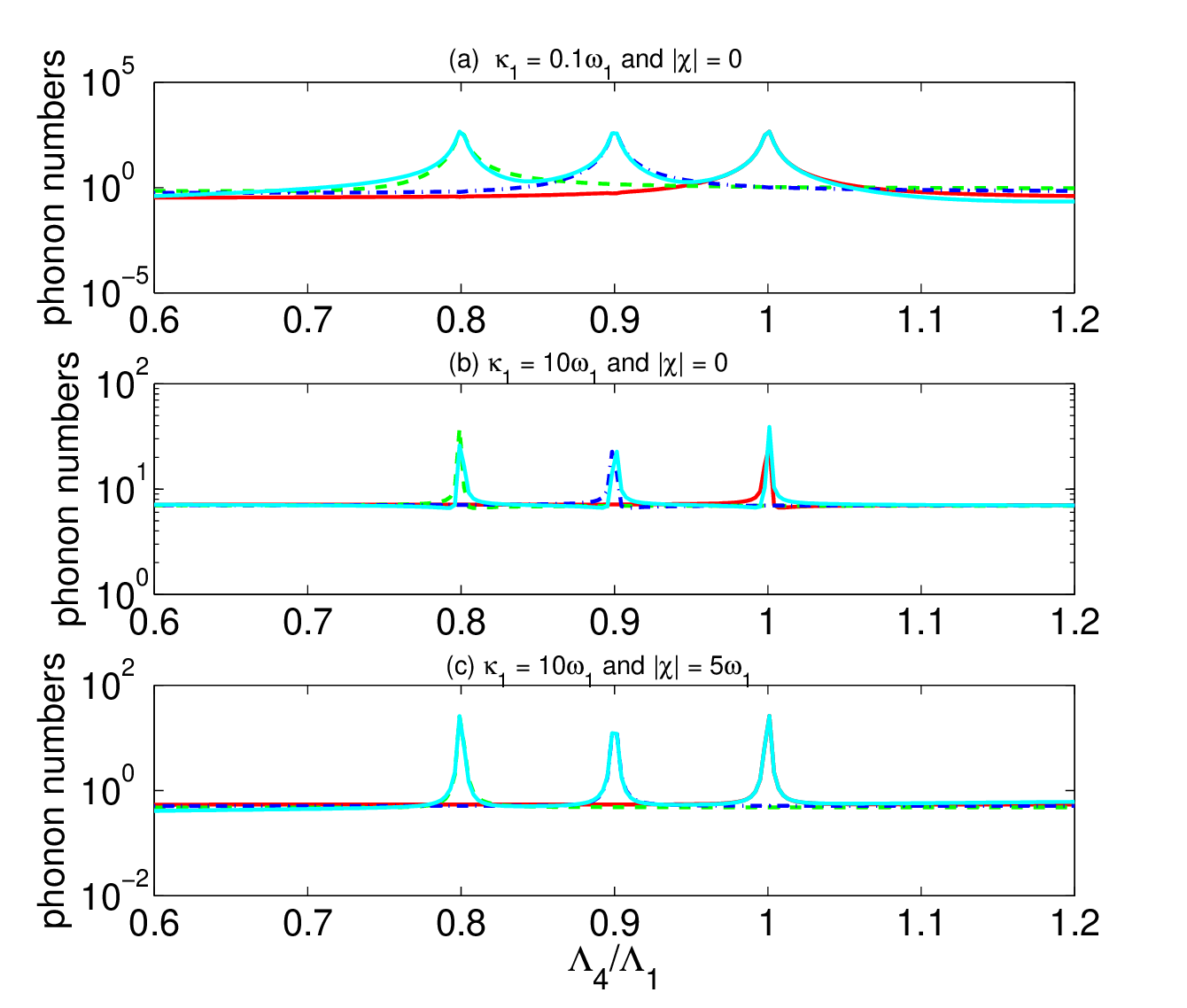}}}
\caption{Steady-state mean phonon numbers $n_1$ (red solid lines), $n_2$ (green dashed lines), $n_3$ (blue dash-dot lines), and $n_4$ (cyan solid lines)  versus $\Lambda_4/\Lambda_1$ for different $\kappa_1$ and $|\chi|$. The parameters are $\gamma_1 = \gamma_2 = \gamma_3 = 10^{-6} \omega_1$, $G_{11} = G_{12} = G_{13} = G_{14} = 0.1\omega_1, \phi = 0.5\pi,
\Delta_c = \omega_1$, $\Lambda_1 = 0.2\omega_1$, $\Lambda_2 = 0.8\Lambda_1$, $\Lambda_3 = 0.9\Lambda_1$, $\omega_1 = \omega_2 = \omega_3 = \omega_4$, and $n_{th} = 1000$.
} \label{fig7}
\end{figure}
Here, we investigate the ground-state cooling of four degenerate mechanical modes.
Similar to the case of two or three mechanical modes, we can find the evolution of the covariance matrix $V$ for four mechanical modes.
The expressions of $A$ and $D$ are not written out explicitly here since they are too long.
So, we only show the numerical results here. In Fig. 7, we plot the steady-state mean phonon numbers $n_1$, $n_2$,
$n_3$, and $n_4$ as functions of $\Lambda_4/\Lambda_1$ for different values of $\kappa_1$ and $|\chi|$
with $\omega_1 = \omega_2 = \omega_3 = \omega_4$. From Fig. 7(a), one can find there are three peaks in the cyan solid lines ($n_4$) which are natural consequences of three possible dark modes formed by mechanical modes. The first peak is located at $\Lambda_4 = 0.8\Lambda_1$ and
the dark mode is mixed by $b_4$ and $b_2$ since $\omega_4 = \omega_2$ and $\Lambda_4 = \Lambda_2$.
In this case, $b_1$ and $b_3$ can be cooled efficiently.
The second peak is located at $\Lambda_4 = 0.9\Lambda_1$ and the dark mode is formed
by $b_4$ and $b_3$ due to $\omega_4 = \omega_3$ and $\Lambda_4 = \Lambda_3$.
The third peak is located at $\Lambda_4 = \Lambda_1$ and the dark mode is formed by $b_4$ and $b_1$.
In the case of $\kappa_1 = 10\omega_1$ and $|\chi| = 0$, all mechanical resonators can not be cooled efficiently [see Fig. 7(b)].
Again, we introduce the optical nonlinearity $\chi$.
As one can clearly see from Fig. 7(c), the simultaneous ground-state cooling of all degenerate mechanical modes can
be realized even in the unresolved sideband regime with the help of
optical nonlinearity and mechanical nonlinearities with different amplitudes.

Finally, we discuss two competing effects of the optical nonlinearity in the present model.
First, the Stokes heating processes can be significantly suppressed and the backaction
limit of standard sideband cooling can be surpassed with the help of the optical nonlinearity as pointed out in Ref. \cite{Asjad2019}.
The first effect of the optical nonlinearity is helpful for ground-state cooling.
Second, there is nonlinearity in the Hamiltonian $H_{\chi} = \frac{i \chi_0}{2} (a_1^{\dag 2} a_2 - a_1^2 a_2^{\dag})$
of Eq. (6) when the optical nonlinearity $\chi_0$ is introduced. This nonlinearity is harmful for ground-state cooling since the
photon number of cavity $a_1$ is increased.
As a result, the influence of the optical nonlinearity on the ground-state cooling of the mechanical modes is a tradeoff between these two competing effects.
In the highly unresolved sideband regime with $\kappa_1 \gg \omega_1$, the photons generated by the optical nonlinearity can decay very
quickly and the first effect which suppresses the Stokes heating processes is dominant.
Thus, the mechanical modes can be cooled efficiently even in the highly unresolved sideband regime.
This is consistent with the results of our work.
In contrast, in the resolved sideband regime with $\kappa_1 \ll \omega_1$, the photons generated by the optical nonlinearity cannot decay quickly.
Although the Stokes heating processes can still be suppressed by the optical nonlinearity in the resolved sideband regime,
these photons generated by the nonlinearity can heat the mechanical modes significantly and is the dominant effect in the resolved sideband regime.
In fact, we observe that the mean phonon numbers of mechanical
modes in the case of $|\chi| > 0 $ can be larger than that of $|\chi| = 0$ in the resolved sideband regime.
The results are not shown here because we focus on the ground-state cooling of mechanical modes in the highly unresolved sideband regime in
the present work. Thus, if we want to cool several degenerate resonators in the resolved sideband regime, the optical nonlinearity
should not be introduced. It is worth noting that one can overcome the dark-mode effect in the resolved sideband regime using other methods
\cite{Lai2018,Lai20212,Lai2020,Naseem2021,Sommer2019,Huang20221,Huang20222}.

\section{Conclusion}
In the present work, we have proposed a scheme to realize simultaneous ground-state cooling of degenerate mechanical
oscillators in the unresolved sideband regime with the help of optical and mechanical nonlinearities.
One main obstacle for ground-state cooling of degenerate mechanical oscillators is the dark-mode effect.
This kind of effect emerges if several degenerate mechanical modes
couple to a common optical mode. In this case the ground-state cooling of mechanical modes is suppressed significantly.
Here, we introduced the Duffing nonlinearities (mechanical nonlinearities)
to break the dark mode formed by degenerate mechanical modes. Our results show that the dark mode can be completely
destroyed by mechanical nonlinearities when their amplitudes are not very close.

In the absence of the second-order nonlinearity medium, the degenerate mechanical resonators can not be cooled efficiently
in the unresolved sideband regime. However, if we put the second-order nonlinearity (optical nonlinearity) into the system,
the simultaneous ground-state cooling of two or more degenerate mechanical modes can be accomplished even in the unresolved sideband regime.

\section*{Acknowledgement}
This project was supported by the National Natural Science Foundation
of China (Grants No. 11775190 and No. 12165007).

\end{document}